\documentclass[11pt]{article}
\usepackage{moriond,epsfig}

\def\be{\begin{equation}}
\def\ee{\end{equation}}
\def\bea{\begin{eqnarray}}
\def\eea{\end{eqnarray}}

\def\gtwid{\mathrel{\raise.3ex\hbox{$>$\kern-.75em\lower1ex\hbox{$\sim$}}}}
\def\ltwid{\mathrel{\raise.3ex\hbox{$<$\kern-.75em\lower1ex\hbox{$\sim$}}}}

\begin{document}
\vspace*{4cm}
\title{THE SEARCH FOR DARK MATTER AXIONS}

\author{ P. SIKIVIE }

\address{Theoretical Physics Division, CERN, CH-1211 Gen\`eve 23, 
Switzerland\\and\\
Department of Physics, University of Florida, Gainesville, FL 32611, USA}

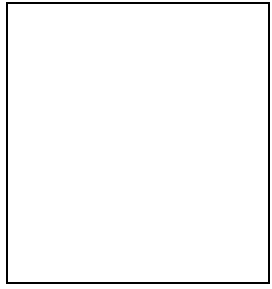
\maketitle\abstracts{
Axions solve the Strong CP Problem and are a cold dark matter 
candidate.  The combined constraints from accelerator searches,
stellar evolution limits and cosmology suggest that the axion 
mass is in the range $3 \cdot 10^{-3} > m_a > 10^{-6}$ eV.  
The lower bound can, however, be relaxed in a number of ways.
I discuss the constraint on axion models from the absence of
isocurvature perturbations.   Dark matter axions can be searched 
for on Earth by stimulating their conversion to microwave photons 
in an electromagnetic cavity permeated by a magnetic field.  Using 
this technique, limits on the local halo density have been obtained
by the Axion Dark Matter eXperiment.}

\section{Introduction}

The standard model of elementary particles has been an extraordinary
breakthrough in our description of the physical world.  It agrees
with experiment and observation disconcertingly well and surely
provides the foundation for all further progress in our field.  It
does however present us with a puzzle.

Indeed the action density includes, in general, a term
\begin{equation}
{\cal L}_{\rm stand~mod} = ...
~+~{\theta g^2\over 32\pi^2} G^a_{\mu\nu} \tilde G^{a\mu\nu}
\label{ggdual}
\end{equation}
where $G^a_{\mu\nu}$ are the QCD field strengths, $g$ is the QCD
coupling constant and $\theta$ is a parameter.  The dots represent
all the other terms in the action density, i.e. the terms that lead
to the numerous successes of the standard model.  Eq.~(\ref{ggdual})
perversely shows the one term that isn't a success.  Using the
statement of the chiral anomaly \cite{abj}, one can show three
things about that term.  First, that QCD physics depends on the
value of the parameter $\theta$ because in the absence of such
dependence QCD would have a $U_A(1)$ symmetry in the chiral limit,
and we know QCD has no such $U_A(1)$ symmetry \cite{SW}.  Second,
that $\theta$ is cyclic, that is to say that physics at $\theta$
is indistinguishable from physics at $\theta + 2\pi$.  Third, that
an overall phase in the quark mass matrix $m_q$ can be removed by
a redefinition of the quark fields only if, at the same time,
$\theta$ is shifted to $\theta - \arg\det m_q$.  The combination of
standard model parameters $\bar{\theta} \equiv \theta - \arg \det m_q$
is independent of quark field redefinitions.  Physics therefore
depends on $\theta$ solely through $\bar{\theta}$.

Since physics depends on $\bar{\theta}$, the value of $\bar{\theta}$ is
determined by experiment.  The term shown in Eq.~(\ref{ggdual}) violates P
and CP.  This source of P and CP violation is incompatible with the
experimental upper bound on the neutron electic dipole moment unless
$|\bar{\theta}| < 10^{-10}$.  A new improved upper limit on the neutron
electric dipole moment ($|d_n| < 3.0 \cdot 10^{-26}~e~$cm) was reported
at this conference by P. Geltenbort \cite{ned}.  The puzzle aforementioned
is why the value of $\bar{\theta}$ is so small.  It is usually referred to
as the ``Strong CP Problem".  If there were only strong interactions, a
zero value of $\bar{\theta}$ could simply be a consequence of P and CP
conservation.  That would not be much of a puzzle. But there are also 
weak interactions and they, and therefore the standard model as a whole,
violate P and CP.  So these symmetries can not be invoked to set
$\bar{\theta} = 0$.  More pointedly, P and CP violation are introduced in
the standard model by letting the elements of the quark mass matrix $m_q$
be arbitrary complex numbers \cite{KM}.  In that case, one expects $\arg
\det m_q$, and hence $\bar{\theta}$, to be a random angle.

The puzzle is removed if the action density is instead
\begin{equation}
{\cal L}_{\rm stand~mod~+~axion} =~...
~+~{1 \over 2}\partial_\mu a \partial^\mu a
~+~{g^2\over 32\pi^2}~{a(x) \over f_a}~G^a_{\mu\nu} \tilde G^{a\mu\nu}
\label{ax}
\end{equation}
where $a(x)$ is a new scalar field, and the dots represent the other
terms of the standard model.  $f_a$ is a constant with dimension of
energy.  The $a G \cdot \tilde G$ interaction in Eq.~(\ref{ax}) is
not renormalizable.  However, there is a recipe for constructing
renormalizable theories whose low energy effective action density
is of the form of Eq.~(\ref{ax}).  The recipe is as follows: construct
the theory in such a way that it has a $U(1)$ symmetry which (1) is a
global symmetry of the classical action density, (2) is broken by the
color anomaly, and (3) is spontaneously broken.  Such a symmetry is
called Peccei-Quinn symmetry after its inventors \cite{PQ}.  Weinberg
and Wilczek \cite{WW} pointed out that a theory with a $U_{\rm PQ}(1)$
symmetry has a light pseudo-scalar particle, called the axion.  The
axion field is $a(x)$.  $f_a$ is of order the expectation value that
breaks $U_{\rm PQ}(1)$, and is called the ``axion decay constant".

In the theory defined by Eq.~(\ref{ax}), $\bar{\theta} = {a(x) \over f_a}
- \det\arg m_q$ depends on the expectation value of $a(x)$.  That
expectation value minimizes the effective potential.  The Strong
CP Problem is then solved because the minimum of the QCD effective
potential $V(\bar{\theta})$ occurs at $\bar{\theta} = 0$ \cite{VW}.
The weak interactions induce a small value for $\bar{\theta}$
\cite{EG,GR}, of order $10^{-17}$, but this is consistent with 
experiment.

The notion of Peccei-Quinn (PQ) symmetry may seem contrived.  Why should
there be a $U(1)$ symmetry which is broken at the quantum level but which
is exact at the classical level?  However, the reason for PQ symmetry may
be deeper than we know at present.  String theory contains many examples
of symmetries which are exact classically but which are broken by
quantum anomalies, including PQ symmetry \cite{Wit,Kiw,Svr}.  Within 
field theory, there are examples of theories with {\it automatic} PQ
symmetry, i.e. where PQ symmetry is a consequence of just the particle
content of the theory without adjustment of parameters to special values.

The first axion models had $f_a$ of order the weak interaction
scale and it was thought that this was an unavoidable property of
axion models.  However, it was soon pointed out \cite{KSVZ,DFSZ} 
that the value of $f_a$ is really arbitrary, that it is possible 
to construct axion models with any value of $f_a$.  A value of 
$f_a$ far from any previously known scale need not lead to a 
hierarchy problem because PQ symmetry can be broken by the 
condensates of a new technicolor-like interaction \cite{Kim}.

The properties of the axion can be derived using the methods of
current algebra \cite{curr}.  The axion mass is given in terms of
$f_a$ by
\begin{equation}
m_a\simeq 6~eV~{10^6 GeV\over f_a}\, .
\label{ma}
\end{equation}
All the axion couplings are inversely proportional to $f_a$.
For example, the axion coupling to two photons is:
\begin{equation}
{\cal L}_{a\gamma\gamma} = -g_\gamma {\alpha\over \pi} {a(x)\over f_a}
\vec E \cdot\vec B~~~\ .
\label{aEB}
\end{equation}
Here $\vec E$ and $\vec B$ are the electric and magnetic fields,
$\alpha$ is the fine structure constant, and $g_\gamma$ is a
model-dependent coefficient of order one.  $g_\gamma=0.36$ in
the DFSZ model \cite{DFSZ} whereas $g_\gamma=-0.97$ in the KSVZ
model \cite{KSVZ}.  The coupling of the axion to a spin 1/2 fermion
$f$ has the form:
\begin{equation}
{\cal L}_{a \overline f f} = i g_f {m_f \over f_a}
a \overline f \gamma_5 f
\label{cf}
\end{equation}
where $g_f$ is a model-dependent coefficient of order one.  In the
KSVZ model the coupling to electrons is zero at tree level.  Models
with this property are called 'hadronic'.


The axion has been searched for in many places but not found \cite{arev}.
The resulting constraints may be summarized as follows.  Axion masses 
larger than about 50 keV are ruled out by particle physics experiments 
(beam dumps and rare decays) and nuclear physics experiments.  The next 
range of axion masses, in decreasing order, is ruled out by stellar 
evolution arguments.  The longevity of red giants rules out
200 keV $> m_a >$ 0.5 eV \cite{astro,Raff87} in the case of hadronic
axions, and 200 keV $> m_a > 10^{-2}$ eV \cite{Schramm} in the case of
axions with a large coupling to electrons [$g_e = 0(1)$ in Eq. \ref{cf}].
The duration of the neutrino pulse from Supernova 1987a rules out
2 eV $> m_a > 3 \cdot 10^{-3}$ eV \cite{1987a}.  Finally, there is a
lower limit, $m_a \gtwid 10^{-6}$ eV, from cosmology which will be
discussed in detail in the next section.  This leaves open an ``axion
window": $3 \cdot 10^{-3} > m_a \gtwid 10^{-6}$ eV.  We will see, 
however, that the lower edge of this window ($10^{-6}$ eV) is much 
softer than its upper edge.

\section{Axion cosmology}

The implications of the existence of an axion for the history of the
early universe may be briefly described as follows.  At a temperature of
order $f_a$, a phase transition occurs in which the $U_{PQ}(1)$ symmetry
becomes spontaneously broken.  This is called the PQ phase transition.
At these temperatures, the non-perturbative QCD effects which produce the
effective potential $V(\overline\theta)$ are negligible \cite{GPY}, the
axion is massless and all values of $\langle a(x)\rangle$ are equally
likely.  Axion strings appear as topological defects.  One must
distinguish two scenarios, depending on wether inflation occurs
with reheat temperature lower (case 1) or higher (case 2) than the
PQ transition temperature.  In case 1 the axion field gets homogenized
by inflation and the axion strings are 'blown' away.

When the temperature approaches the QCD scale, the potential
$V(\overline\theta)$ turns on and the axion acquires mass.  There is
a critical time, defined by $m_a(t_1)t_1 = 1$, when the axion field
starts to oscillate in response to the turn-on of the axion mass.
The corresponding temperature $T_1 \simeq 1$ GeV \cite{ac}.  The
initial amplitude of this oscillation is how far from zero the axion
field lies when the axion mass turns on.  The axion field oscillations
do not dissipate into other forms of energy and hence contribute to the
cosmological energy density today \cite{ac}. This contribution is called
of `vacuum realignment'.  It is further described below.  Note that the
vacuum realignment contribution may be accidentally suppressed in case 1
if the homogenized axion field happens to lie close to zero.

In case 2 the axion strings radiate axions \cite{rd,Har} from the
time of the PQ transition till $t_1$ when the axion mass turns on.   At
$t_1$ each string becomes the boundary of $N$ domain walls.  If $N=1$,
the network of walls bounded by strings is unstable \cite{Ev,Paris} and
decays away.  If $N>1$ there is a domain wall problem \cite{adw} because
axion domain walls end up dominating the energy density, resulting in a
universe very different from the one observed today.  There is a way
to avoid this problem by introducing an interaction which slightly
lowers one of the $N$ vacua with respect to the others.  In that
case, the lowest vacuum takes over after some time and the domain walls
disappear.  There is little room in parameter space for that to happen
and we will not consider this possibility \cite{axwall} further here.  
Henceforth, we assume $N=1$.

In case 2 there are three contributions to the axion cosmological
energy density.  One contribution \cite{rd,Har,thA,Hag,Shel,Yam,us}
is from axions that were radiated by axion strings before $t_1$.  A
second contribution is from axions that were produced in the decay
of walls bounded by strings after $t_1$ \cite{Hag,Ly,Nag,axwall}.  A
third contribution is from vacuum realignment \cite{ac}.

Let me briefly indicate how the vacuum alignment contribution is
evaluated.  Before time $t_1$, the axion field did not oscillate
even once.  Soon after $t_1$, the axion mass is assumed to change
sufficiently slowly that the total number of axions in the
oscillations of the axion field is an adiabatic invariant.  The
average number density of axions at time $t_1$ is
\begin{equation}
n_a(t_1)\simeq {1\over 2} m_a(t_1) \langle a^2(t_1)\rangle \simeq
\pi f_a^2 {1\over t_1}
\label{nat1}
\end{equation}
In Eq.~(\ref{nat1}), we used the fact that the axion field $a(x)$
is approximately homogeneous on the horizon scale $t_1$.  Wiggles
in $a(x)$ which entered the horizon long before $t_1$ have been
red-shifted away \cite{Vil}.  We also used the fact that the initial
departure of $a(x)$ from the nearest minimum is of order $f_a$.  The
axions of Eq.~(\ref{nat1}) are decoupled and non-relativistic.
Assuming that the ratio of the axion number density to the
entropy density is constant from time $t_1$ till today, one
finds \cite{ac,axwall}
\begin{equation}
\Omega_a \simeq {1 \over 2}
\left({f_a \over 10^{12}{\rm GeV}}\right)^{7\over 6}
\left({0.7 \over h}\right)^2
\label{oma}
\end{equation}
for the ratio of the axion energy density to the critical density
for closing the universe.  $h$ is the present Hubble rate in units
of 100 km/s.Mpc.  The requirement that axions do not overclose the
universe implies the constraint $m_a \gtwid 6 \cdot 10^{-6}$~ eV.

The contribution from axion string decay has been debated over the
years.  The main issue is the energy spectrum of axions radiated
by axion strings.  Battye and Shellard \cite{Shel} have carried out
computer simulations of bent strings (i.e. of wiggles on otherwise
straight strings) and have concluded that the contribution from
string decay is approximately ten times larger than that from vacuum
realignment, implying a bound on the axion mass approximately ten
times more severe, say $m_a \gtwid 6 \cdot 10^{-5}$ eV instead of
$m_a \gtwid 6 \cdot 10^{-6}$ eV.  My collaborators and I have done
simulations of bent strings \cite{Hag}, of circular string loops
\cite{Hag,us} and non-circular string loops \cite{us}.  We conclude
that the string decay contribution is of the same order of magnitude
than that from vacuum realignment.  Yamaguchi, Kawasaki and Yokoyama
\cite{Yam} have done computer simulations of a network of strings in
an expanding universe, and concluded that the contribution from string
decay is approximately three times that of vacuum realignment.  The
contribution from wall decay has been discussed in detail in
ref.~\cite{axwall}.  It is probably subdominant compared to the
vacuum realignment and string decay constributions.

It should be emphasized that there are many sources of uncertainty in the
cosmological axion energy density aside from the uncertainty about the
contribution from string decay.  The axion energy density may be diluted
by the entropy release from heavy particles which decouple before the QCD
epoch but decay afterwards \cite{ST}, or by the entropy release associated
with a first order QCD phase transition.  On the other hand, if the QCD
phase transition is first order \cite{pt}, an abrupt change of the axion
mass at the transition may increase $\Omega_a$. If inflation occurs with
reheat temperature less than $T_{PQ}$, there may be an accidental
suppression of $\Omega_a$ because the homogenized axion field happens to
lie close to a $CP$ conserving minimum.  Because the RHS of Eq.~(7) is
multiplied in this case by a factor of order the square of the initial
vacuum misalignment angle ${a(t_1)\over f_a}$ which is randomly chosen
between $-\pi$ and $+\pi$, the probability that $\Omega_a$ is suppressed
by a factor $x$ is of order $\sqrt{x}$.  Recently, Kaplan and Zurek
proposed a model \cite{KZ} in which the axion decay constant $f_a$ is
time-dependent, the value $f_a(t_1)$ during the QCD phase-transition 
being much smaller than the value $f_a$ today. This yields a suppression
of the axion cosmological energy density by a factor
$({f_a(t_1) \over f_a})^2$ compared to the usual case [replace $f_a$ by
$f_a(t_1)$ in Eq.~(\ref{nat1})].  Finally, the axion density may be
diluted by 'coherent deexcitation', i.e. adiabatic level crossing of
$m_a(t)$ with the mass of some other pseudo-Nambu-Goldstone boson which 
mixes with the axion \cite{HR}.

The axions produced when the axion mass turns on during the QCD phase
transition are cold dark matter (CDM) because they are non-relativistic
from the moment of their first appearance at 1~GeV temperature.  Studies
of large scale structure formation support the view that the dominant
fraction of dark matter is CDM.  Any form of CDM necessarily contributes
to galactic halos by falling into the gravitational wells of galaxies.
Hence, there is excellent motivation to look for axions as constituent
particles of our galactic halo.

There is a particular kind of clumpiness \cite{amc,axwall} which affects
axion dark matter if there is no inflation after the Peccei-Quinn phase
transition (case 2).  This is due to the fact that the dark matter axions
are inhomogeneous with $\delta \rho / \rho \sim 1$ over the horizon scale
at temperature $T_1 \simeq$ 1 GeV, when they are produced at the start of
the QCD phase-transition, combined with the fact that their velocities are
so small that they do not erase these inhomogeneities by free-streaming
before the time $t_{eq}$ of equality between the matter and radiation
energy densities when matter perturbations can start to grow.  These
particular inhomogeneities in the axion dark matter are in the non-linear
regime immediately after time $t_{eq}$ and thus form clumps, called `axion
mini-clusters' \cite{amc}.  They have mass $M_{mc} \simeq 10^{-13}
M_\odot$ and size $l_{mc} \simeq 10^{13}$ cm.

\section{Axion isocurvature perturbations}

If inflation occurs after the Peccei-Quinn phase transition, i.e. if 
the reheat temperature after inflation $T_{\rm RH}$ is less than the 
temperature $T_{\rm PQ}$ at which $U_{\rm PQ}(1)$ is restored (case 1), 
the quantum mechanical fluctuations of the axion field during the
inflationary epoch cause isocurvature density perturbations 
\cite{iso,Turn91} in the early universe.  The cosmic microwave 
background obervations are consistent with purely adiabatic 
density perturbations and therefore place a constraint, which
we now discuss.

Fluctuations generated during inflation in a massless weakly coupled 
scalar field, such as the inflaton or the axion, are characterized 
by the power spectrum \cite{qfl} 
\begin{equation}
P_a (k) \equiv \int {d^3 x \over (2 \pi)^3}
<\delta a(\vec{x},t) \delta a(\vec{x}^\prime,t)>
e^{- i \vec{k} \cdot (\vec{x} - \vec{x}^\prime)} =
\left({H_I \over 2 \pi}\right)^2 {2 \pi^2 \over k^3}~~~\ ,
\label{axpow}
\end{equation}
where $\vec{x}$ are comoving coordinates.  Eq.~(\ref{axpow}) is
often written in the shorthand notation $\delta a = {H_I \over 2 \pi}$.  
The axion field fluctuations are ``frozen" from the time their wavelengths 
exceed the horizon size $H_I^{-1}$ during the inflationary epoch till
the time their wavelenghs reenter the horizon long after inflation has 
ended.

At the start of the QCD phase transition, the local value of the
axion field $a(\vec{x},t)$ determines the local number density of
cold axions produced by the vacuum realignment mechanism 
\begin{equation}
n_a(\vec{x}, t_1) = {f_a^2 \over 2 t_1} \alpha(\vec{x}, t_1)^2
\label{locna1}
\end{equation}
where $\alpha(\vec{x}, t_1) = a(\vec{x}, t_1)/f_a$ is the local
misalignment angle.  The fluctuations in the axion field produce
perturbations in the axion dark matter density
\begin{equation}
{\delta n_a^{\rm iso} \over n_a} = {2 \delta a \over a_1} =
{H_I \over \pi f_a \alpha_1}
\label{deltana}
\end{equation}
where $a_1 = a(t_1) = f_a \alpha_1$ is the initial value of the
axion field, at the start of the QCD phase transition, common
to our entire visible universe.  These perturbations initially obey
$\delta \rho_a^{\rm iso} = - \delta \rho_{\rm r}^{\rm iso}$ since
the vacuum realignment mechanism converts energy stored in the
quark-gluon plasma into axion rest mass energy.  Such
perturbations are commonly called ``isocurvature perturbations"
because they do not initially produce a source for the Newtonian
potential $\Phi = {1 \over 2}(g_{00} - 1)$. Note that in case 1 
the density perturbations in the cold axion fluid have both adiabatic 
and isocurvature components.  The adiabatic perturbations
(${\delta \rho_a^{\rm ad} \over 3 \rho_a} =
{\delta \rho_{\rm r}^{\rm iso} \over 4 \rho_{\rm r}}
= {\delta T \over T}$) are produced by the quantum
mechanical fluctuations of the inflaton field during
inflation, whereas the isocurvature perturbations
[$\delta \rho_a^{\rm iso}(t_1) \simeq - \delta \rho_r(t_1)^{\rm iso}$]
are produced by the quantum mechanical fluctuations of
the axion field during that same epoch.
 
Isocurvature perturbations make a different imprint on the
cosmic microwave background than do adiabatic ones.  The
CMBR observations are consistent with pure adiabatic
perturbations.  According to P. Crotty et al. \cite{Crot03},
the fraction of cold dark matter perturbations which are
isocurvature can not be larger than 31\%.  This places
a constraint on axion models if the Peccei-Quinn phase
transition occurs before inflation (case 1).  Allowing for 
the possibility that only part of the cold dark matter is 
axions, the CMBR constraint of ref. \cite{Crot03} is
\begin{equation}
{\delta \rho_a^{\rm iso} \over \rho_{\rm CDM}} =
{\delta\rho_a^{\rm iso} \over \rho_a} \cdot {\rho_a \over \rho_{\rm CDM}}
= {H_I \over \pi f_a \alpha_1} {\Omega_a \over \Omega_{\rm CDM}} <
0.31~{\delta \rho_m \over \rho_m}~~~~\ ,
\label{isocon}
\end{equation}
where we used Eq.~(\ref{deltana}). ${\delta \rho_m \over \rho_m}$
is the amplitude of the primordial spectrum of matter perturbations.
It is related to the amplitude of large scale (low multipole)
CMBR anisotropies through the Sachs-Wolfe effect
\cite{Sach}.  The observations imply \cite{Dode03}
${\delta \rho_m \over \rho_m} \simeq 4.6~10^{-5}$.

In terms of $\alpha_1$, the cold axion energy density is given
in case 1 by 
\begin{equation}
\Omega_a \simeq 0.15
\left({f_a \over 10^{12} {\rm GeV}}\right)^{7 \over 6}\alpha_1^2~~~\ .
\label{case1}
\end{equation}
where we assumed $h \simeq 0.7$.  It has been remarked by many authors, 
starting with S.-Y. Pi \cite{Pi84}, that it is possible for $f_a$ to be 
much larger than $10^{12} {\rm GeV}$ because $\alpha_1$ may be accidentally
small in our visible universe.  The requirement that 
$\Omega_a < \Omega_{\rm CDM} = 0.22$ \cite{WMAP3} implies
\begin{equation}
|{\alpha_1 \over \pi}| < 0.5
\left({10^{12} {\rm GeV} \over f_a}\right)^{7 \over 12}~~~\ .
\label{alpha1}
\end{equation}
Since $- \pi < \alpha_1 < + \pi$ is the a-priori range of
$\alpha_1$ values and no particular value is preferred over
any other, $|{\alpha_1 \over \pi}|$ may be taken to be the
``probability" that the initial misalignment angle has
magnitude less than $|\alpha_1|$.  If
$|{\alpha_1 \over \pi}| = 4 \cdot 10^{-3}$, for example,
$f_a$ may be as large as $10^{16}$ GeV.

The presence of isocurvature perturbations constrains the
small $\alpha_1$ scenario in two ways \cite{Turn91}.  First,
it makes it impossible to have $\alpha_1$ arbitrarily small
since
\begin{equation}
\alpha_1 > \delta \alpha_1 = {H_I \over 2 \pi f_a}~~~\ .
\label{minalpha1}
\end{equation}
Combining Eqs.~(\ref{alpha1}) and (\ref{minalpha1}), we
obtain the bound
\begin{equation}
\Lambda_I < 6 \cdot 10^{15} {\rm GeV}
\left({f_a \over 10^{12} {\rm GeV}}\right)^{5 \over 24}~~~\ .
\label{con1}
\end{equation}
Second, one must require axion isocurvature perturbations to be
consistent with CMBR observations.  Combining Eqs.~(\ref{isocon})
and (\ref{case1}), and setting $\Omega_{\rm CDM} = 0.22$,
${\delta \rho_m \over \rho_m} = 4.6~10^{-5}$, we obtain
\begin{equation}
\Lambda_I < 10^{13} {\rm GeV}~~\Omega_a^{-{1 \over 4}}~
\left({f_a \over 10^{12} {\rm GeV}}\right)^{5 \over 24}~~~\ .
\label{con2}
\end{equation}
Let us keep in mind that the bounds (\ref{con1}) and (\ref{con2})
pertain only if $T_{\rm RH} < T_{\rm PQ}$.  One may, for
example, have $\Omega_a = 0.22$, $f_a \simeq 10^{12}$ GeV,
and $\Lambda_I \simeq 10^{16}$ GeV, provided
$T_{\rm RH} \gtwid 10^{12}$ GeV, which is possible if reheating
is sufficiently efficient.

\section{Dark matter axion detection}

An electromagnetic cavity permeated by a strong static magnetic field
can be used to detect galactic halo axions \cite{ps}.  The relevant
coupling is given in Eq.~(\ref{aEB}). Galactic halo axions have
velocities $\beta$ of order $10^{-3}$ and hence their energies
$E_a=m_a+{1\over 2} m_a\beta^2$ have a spread of order $10^{-6}$
above the axion mass.  When the frequency $\omega=2\pi f$ of a
cavity mode equals $m_a$, galactic halo axions convert resonantly
into quanta of excitation (photons) of that cavity mode.  The power
from axion $\to$ photon conversion on resonance is found to
be \cite{ps,kal}:
\begin{eqnarray}
P&=&\left ({\alpha\over\pi} {g_\gamma\over f_a}\right )^2 V\, B_0^2
\rho_a C {1\over m_a} \hbox{Min}(Q_L,Q_a)~~~~\nonumber\\
&=& 0.5\; 10^{-26} \hbox{Watt}\left( {V\over 500\hbox{\ liter}}\right)
\left({B_0\over 7\hbox{\ Tesla}}\right)^2 
~C \left({g_\gamma \over 0.36}\right)^2~~~\cdot\nonumber\\
~~~&\cdot&\left({\rho_a\over {1\over 2} \cdot 10^{-24}
{{\rm gr} \over \hbox{\rm cm}^3}}\right) 
~\left({m_a\over 2\pi (\hbox{GHz})}\right)\hbox{Min}(Q_L,Q_a)
\end{eqnarray}
where $V$ is the volume of the cavity, $B_0$ is the magnetic field
strength, $Q_L$ is its loaded quality factor, $Q_a=10^6$ is the
`quality factor' of the galactic halo axion signal (i.e. the ratio of
their energy to their energy spread), $\rho_a$ is the density of
galactic halo axions on Earth, and $C$ is a mode dependent form factor
given by
\begin{equation}
C = {\left| \int_V d^3 x \vec E_\omega \cdot \vec B_0\right|^2
\over B_0^2 V \int_V d^3x \epsilon |\vec E_\omega|^2}  \,
\end{equation}
where $\vec B_0(\vec x)$ is the static magnetic field,
$\vec E_\omega(\vec x) e^{i\omega t}$ is the oscillating electric
field and $\epsilon$ is the dielectric constant.  Because the axion
mass is only known in order of magnitude at best, the cavity must be
tunable and a large range of frequencies must be explored seeking a
signal.  The cavity can be tuned by moving a dielectric rod or metal
post inside it.

For a cylindrical cavity and a homogeneous longitudinal magnetic field,
$C=0.69$ for the lowest TM mode.  The form factors of the other modes
are much smaller.  The resonant frequency of the lowest TM mode of a
cylindrical cavity is $f$=115 MHz $\left( {1m\over R}\right)$ where $R$
is the radius of the cavity.  Since $10^{-6}\hbox{\ eV} = 2\pi$ (242 MHz),
a large cylindrical cavity is convenient for searching the low frequency
end of the range of interest.  To extend the search to high frequencies
without sacrifice in volume, one may power-combine many identical cavities
which fill up the available volume inside a magnet's bore \cite{rsci,hag}.
This method allows one to maintain $C=0(1)$ at high frequencies, albeit
at the cost of increasing engineering complexity as the number of
cavities increases.


Axion dark matter searches were carried out at Brookhaven National
Laboratory \cite{RBF}, the University of Florida \cite{UF}, Kyoto
University \cite{Kyoto}, and by the ADMX collaboration
\cite{PRL,PRD,ApJL,PRDRC,Duf} at Lawrence Livermore National
Laboratory. Thus far, the ADMX experiment has ruled out 
KSVZ coupled axions at the nominal halo density of \cite{Gates}
$\rho_a = 7.5~10^{-25}~$g/cm$^3$ over the mass range
$1.90 < m_a < 3.35~\mu$eV \cite{ApJL,PRDRC}. The ADMX experiment is
presently being upgraded to replace the HEMT (high electron mobility
transistors) receivers we have used so far with SQUID microwave
amplifiers. HEMT receivers have noise temperature $T_n \sim 3~K$
\cite{Bradley} whereas $T_n \sim 0.05~K$ was achieved with SQUIDs
\cite{Clarke}.  In a second phase of the upgrade, the experiment 
will be equipped with a dilution refrigerator to take full advantage 
of the lowered electronic noise temperature.  When both phases of 
the upgrade are completed, the ADMX detector will have sufficient 
sensitivity to detect axions at even a fraction of the halo density.

The ADMX experiment is equipped with a high resolution spectrometer
which allows us to look for narrow peaks in the spectrum of microwave
photons caused by discrete flows, or streams, of dark matter axions in
our neighborhood. In many discussions of cold dark matter detection
it is assumed that the distribution of CDM particles in galactic halos
is isothermal. However, there are excellent reasons to believe that a
large fraction of the local density of cold dark matter particles is
in discrete flows with definite velocities \cite{is}.  Indeed, because
CDM has very low primordial velocity dispersion and negligible
interactions other than gravity, the CDM particles lie on a 3-dim.
hypersurface in 6-dim.  phase-space.  This implies that the velocity
spectrum of CDM particles at any physical location is discrete, i.e.,
it is the sum of distinct flows each with its own density and velocity.

We searched for the peaks in the spectrum of microwave photons from
axion to photon conversion that such discrete flows would cause in
the ADMX detector.  We found none and placed limits \cite{Duf} on
the density  of any local flow of axions as a function of the flow
velocity dispersion over the axion mass range 1.98 to 2.17 $\mu$eV.
Our limit on the density of discrete flows is approximately a factor
three more severe than our limit on the total local axion dark matter
density.

\section*{Acknowledgments}
This work is supported in part by the U.S. Department of Energy 
under grant DE-FG02-97ER41029.  I am grateful to R. Woodard and 
A. Riotto for helpful discussions on topics related to axion 
isocurvature perturbations.

\end{document}